\documentclass[aps,prd,onecolumn,superscriptaddress,showpacs,floatfix,10pt]{revtex4}
\usepackage{graphicx}% Include figure files
\usepackage{amssymb,amsmath}
\usepackage{subfigure}
\usepackage{epsfig}

\begin{document}

\title{Matter localization and resonant deconfinement in a two-sheeted spacetime}

%\author{Micha\"{e}l Sarrazin
%\address{Laboratoire de Physique du Solide, \\Facult\'es
%Universitaires Notre-Dame de la Paix, \\61 rue de Bruxelles,
%B-5000 Namur, Belgium\\michael.sarrazin@fundp.ac.be} \and Fabrice
%Petit
%\address{Belgian Ceramic Research Centre,\\4 avenue du gouverneur
%Cornez, B-7000 Mons, Belgium\\f.petit@bcrc.be} }

\author{Micha\"{e}l Sarrazin}
\email{michael.sarrazin@fundp.ac.be} \affiliation{Laboratoire de
Physique du Solide, Facult\'es Universitaires Notre-Dame de la Paix,
\\61 rue de Bruxelles, B-5000 Namur, Belgium}

\author{Fabrice Petit}
\email{f.petit@bcrc.be} \affiliation{Belgian Ceramic Research
Centre,\\4 avenue du gouverneur Cornez, B-7000 Mons, Belgium}

\begin{abstract}
In recent papers, a model of a two-sheeted spacetime $M_{4} \times
Z_{2}$ was introduced and the quantum dynamics of massive fermions
was studied in this framework. In the present study, we show that
the physical predictions of the model are perfectly consistent with
observations and most important, it can solve the puzzling problem
of the four-dimensional localization of the fermion species in
multidimensional spacetimes. It is demonstrated that fermion
localization on the sheets arises from the combination of the
discrete bulk structure and environmental interactions. The
mechanism described in this paper can be seen as an alternative to
the domain wall localization arising in continuous five dimensional
spacetimes. Although tightly constrained, motions between the sheets
are, however, not completely prohibited. As an illustration, a
resonant mechanism through which fermion oscillations between the
sheets might occur is described.
\end{abstract}

\pacs{11.10.Kk, 11.25.Wx, 13.40.-f}

\maketitle

\section{Introduction}

Extradimensions are the backbone of present theoretical physics.
During the last years, there has been a considerable interest in
Kaluza-Klein like scenarios suggesting that our usual spacetime
could be just a slice of a larger dimensional manifold. Recent
advances in string theory have thus postulated the existence of
"braneworlds" ($3$-hyperdimensional surfaces embedded in a $N>4$
dimensional manifold) in which we are living on [1,2]. Several
issues like the hierarchy between the electroweak and the Planck
scales [3,4] as well as the idea that the post-inflationary epoch of
our universe was preceded by the collision of D3-branes [5], for
instance, have been successfully addressed in such multidimensional
approaches. \\In braneworld models, it is generally assumed that the
standard model particles can not freely propagate into unseen
dimensions and must be constrained to live on a $3+1$ submanifold
[1,2]. Such conservative point of view adopts the idea that no
interaction (except gravitation) can propagate into the bulk or
between adjacent branes [6,7,8]. The problem is then to find a
physically reasonable mechanism that could effectively constrain the
particles to stay confined in a lower dimensional spacetime sheet
[1,2,9-20]. In some string theory inspired models, the confinement
arises as a natural consequence of the fact that particles are open
strings whose endpoints are attached on Dp-branes [1,2]. Other
approaches postulate the existence of domain walls in which standard
model particles could be entrapped [9-20]. Among earlier works
[9,10,11], Rubakov and Shaposhnikov [10] suggested long time ago
that the wave function of fermion zero modes could concentrate near
domain walls, generating 4D massless chiral fermions attached to
them. This idea has attracted much interest over the last years as
it can solve in a very elegant way the hierarchy problem [3,4] and
the proton decay problem : the extradimensional separation between
chiral fermions generates exponentially suppressed couplings between
them [21-23]. Nevertheless, most of these scenarios suffer from a
significant drawback as they require the existence of external
non-gravitational forces like scalar fields. Since the existence of
such fields is still the subject of debate [24], the credibility of
these scenarios can be reasonably questioned.\\ In parallel, some
attempts have tried to extend the hypothetical graviton capability
of moving through the bulk [1,6,7,8] to the case of massive
particles as well [12,15,16]. As a result, there could be a
possibility for highly massive or energetic particles to acquire a
non zero extradimensional momentum and escape from the branes to
propagate into the bulk. \\Considering these two opposite
approaches, it is clear that the question of whether standard model
particles are totally or only partially confined on 4D hypersurfaces
is still an open issue. \\Furthermore, all those approaches
perpetuate the tradition inherited from relativity by assuming that
the whole universe behaves like a smooth continuum. However, there
have been some recent attempts to develop models where the
continuous extra dimensions are substituted by discrete dimensions
[25-29]. In those approaches, the extra-dimensions are replaced by a
finite number of points and the whole universe can be seen as made
up of a collection of 4D sheets. Besides keeping the physical
richness of multidimensional spacetimes, such multi-sheeted
approaches provide also a nice framework where the standard model
may arise from pure geometry [25]. In recent works, at the crossroad
between brane models and non commutative two-sheeted spacetimes,
present authors have studied the quantum dynamics of fermions in a
$M_{4} \times Z_{2}$ universe [30-32]. Such a model can be seen as a
formal extension of the Kaluza-Klein spacetime with a fifth
extradimension restricted to only two points. It was emphasised that
despite the discrete bulk structure, both spacetime sheets are still
connected at the quantum level. This connection was shown to be
related to the specific geometrical structure of the bulk with a
coupling strength
partly dictated by the magnitude of the electromagnetic gauge fields of both sheets.\\
In this paper, we propose to discuss the predictions of our model
for the issue concerning the fermion localization on the 4D
sub-manifolds. In section 2 we briefly summarize the mathematics
underlying our approach and the main results of previous works are
reported. In section 3, we detail the mechanisms by which fermions
are effectively confined on the 4D sheets. It is demonstrated that
the confining effect arises from the combination of the discrete
bulk structure and the environmental interactions. To the authors
knowledge, the mechanism described here is completely new and
presents the advantage of not relying neither on any exotic trapping
mechanism involving negative bulk energy nor on any scalar fields.\\
Besides, in section 4, we show that particles can have access
granted to the other spacetime sheet in some specific circumstances.
This situation which is analogous to a motion through the bulk in
braneworld models is illustrated by considering a resonant mechanism
forcing the particle to oscillate between the two sheets. This
mechanism which might be experimentally investigated at present
times and energy scales is worth being studied as it could become a
practical tool for the experimental search of extradimensions.

\section{Physical and mathematical framework}

In Refs. [30-32], a model aiming at describing the quantum dynamics
of fermions in a two-sheeted spacetime was described. It corresponds
formally to the product of a four continuous manifold times a
discrete two points space, i.e. $X=M_4\times Z_2$. \\Such a universe
can be pictured as a five dimensional universe where the
fifth dimension is replaced by a discrete two points space with coordinates $%
\pm \delta /2$. Each point is then endowed with its own
four-dimensional sheet and both sheets are separated by a distance
$\delta$. At that point, it is important to stress that this
distance is just a phenomenological one and does not necessarily
match the concept of distance we are familiar with. The interaction
between both sheets is reflected by the existence of a coupling
strength $g=1/\delta$ (see figure 1) given by the inverse distance
between the sheets. In the abovementioned references, it was
demonstrated that the model could be built by considering two
different approaches which are now briefly described.\\

\begin{figure}[tbp]
\centerline{\ \includegraphics[width=7 cm, height=4 cm]{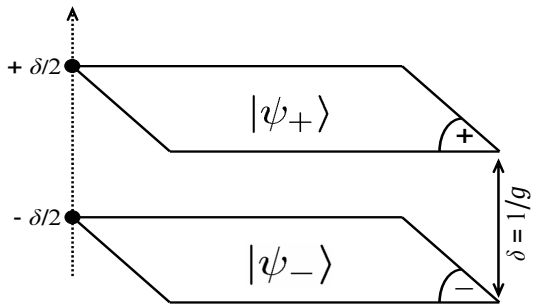}}
\caption{Naive view of the two-sheeted spacetime under
consideration. The coupling constant $g$ between the sheets is
related to the distance which separates them in the extra-space.}
\label{fig1}
\end{figure}

The first approach was mainly based on the work of Connes [25], Viet
and Wali [26,27] and relies on a non-commutative definition of the
exterior derivative acting on the product manifold. Due to the
specific geometrical structure of the bulk, this operator is given
by

\begin{equation}
D_\mu =\left( {{\
\begin{array}{cc}
{\partial _\mu } & 0 \\
0 & {\partial _\mu }
\end{array}
}}\right) ,\text{ }\mu =0,1,2,3\text{ and\ }D_5=\left( {{\
\begin{array}{cc}
0 & g \\
{-g} & 0
\end{array}
}}\right)  \label{1}
\end{equation}
Where the term $g$ has the dimension of mass (in $\hbar = c = 1$
units) and acts as a finite difference operator along the discrete
dimension. In the model discussed in Ref. [30] and contrary to
previous works, $g$ was considered as a constant geometrical field
(and not the Higgs field [25]). As a consequence a mass term was
introduced and a two-sheeted Dirac equation was derived

\begin{eqnarray}
\left( {iD\!\!\!/ -M}\right) \Psi = \left( {i\Gamma
^ND_N-M}\right)\Psi = \left( {{\
\begin{array}{cc}
{i\gamma ^\mu \partial _\mu -m} & {ig\gamma ^5} \\
{ig\gamma ^5} & {i\gamma ^\mu \partial _\mu -m}
\end{array}
}}\right) \left( {{\
\begin{array}{c}
{\psi _{+}} \\
{\psi _{-}}
\end{array}
}}\right)= 0  \label{2}
\end{eqnarray}

It can be noticed that by virtue of the two-sheeted structure of
spacetime, the wave function $\psi$ of the fermion is split into two
components, each component living on a distinct spacetime sheet.
Obviously, the equation (2) can also be derived from the $M_{4}
\times Z_{2}$ lagrangian \textsl{L} defined as
\begin{equation}
\textsl{L}=\bar{\Psi} \left( {i\not{D}-M}\right) \Psi \label{3}
\end{equation}
Where $\bar{\Psi}=\left( {\bar{\psi}_{+},\bar{\psi}_{-}}\right) $ is
the
two-sheeted adjoint spinor of $\Psi$ and with $\bar{\psi}_{+}$ and $\bar{%
\psi}_{-}$ the adjoint spinors respectively of $\psi _{+}$ and $\psi
_{-}$. Such a lagrangian can be also written using the following
expended form
\begin{eqnarray}
\textsl{L} &=&  \bar{\psi}_{+} \left( {i\not{\partial}-m}\right)
\psi _{+} + \bar{\psi}_{-} \left( {i\not{\partial}-m}\right) \psi
_{-} + ig \bar{\psi}_{+}\gamma ^5\psi _{-} + ig\bar{\psi}_{-}\gamma
^5\psi _{+} \label{4}
\end{eqnarray}
At first sight, the doubling of the wave function can be seen as a
reminiscence of the hidden-sector concept. While it is true that
hidden sector models and present approach share several common
points, it is equally true that they differ by the number of
spacetime sheets they consider. For instance, the so-called mirror
matter approach, considers only one 4D manifold and justifies for
the left/right parity by introducing new internal degrees of freedom
to particles (see for instance Ref. [33]). In the present work, it
can be noticed that the number of particle families remains
unchanged but the particles, thanks to their five dimensional
nature, have access granted to the two 4D sheets. So, instead of
considering new internal degrees of freedom, we simply allow the
particles to move freely in an extended spacetime. \\The second
approach developed in Ref. [31] started from the usual covariant
Dirac equation in 5D. By assuming a discrete structure of the bulk,
with two four-dimensional submanifolds, the extradimensional
derivative has to be replaced by a finite difference counterpart
\begin{equation}
(\partial _5\psi )_{\pm }=\pm g(\psi _{+}-\psi _{-})  \label{5}
\end{equation}
Then, as in the non-commutative approach, the Dirac equation breaks
down into a set of two coupled differential equations similar to Eq.
(2) [31]. \\The incorporation of the electromagnetic field in the
model can be done quite easily [30-32]. The usual U(1) gauge field
must be substituted by an extended U(1)$\bigotimes$U(1) gauge
relevant for the discrete $Z_2$ structure of the universe. In
addition, each sheet possesses its own current and charge density
distribution as source of the electromagnetic field. The most
general form for the new gauge (to be incorporated within the
two-sheeted Dirac equation such that $D\!\!\!/\rightarrow
D\!\!\!/+A\!\!\!/$) is then defined by (see also Refs. [28] and [29]
where such a gauge was also considered)

\begin{equation}
A\!\!\!/=\left( {{\
\begin{array}{cc}
{iq\gamma ^\mu A_{+,\mu }} & {\gamma ^5\chi } \\
{\gamma ^5\chi ^{\dagger }} & {iq\gamma ^\mu A_{-,\mu }}
\end{array}
}}\right)  \label{6}
\end{equation}

On the two sheets live then the distinct $\mathbf{A}_{+}$ and $\mathbf{A}%
_{-} $ fields. In the present model, the component $\chi$ cannot be
associated with the usual Higgs field encountered in GSW model.
$\chi$ is the coupling of the photons fields of the two sheets.
Since this term
introduces obvious complications unless to be weak enough compared to $%
\mathbf{A}_{\pm}$ (most notably it leads to unusual transformations
laws of the electromagnetic field which are difficult to reconcile
with observations) it is preferable to set $\chi=0$. The
electromagnetic fields of both sheets are then completely decoupled
and each sheet is endowed with
its own electromagnetic structure. Note that the consequence of having $%
\chi\neq 0$ would have been to couple each charged particle with the
electromagnetic field of both sheets, irrespective of the
localization of this particle in the bulk. For instance a particle
of charge $q$ localized in the sheet $(+)$ would have been sensitive
to the electromagnetic field of
the sheet $(-)$ with an effective charge $\varepsilon q$ ($\varepsilon < 1$%
). This kind of exotic interactions has been considered previously
in literature within the framework of the mirror matter paradigm and
is not covered by the present paper [34,35]. Another relevant
consequence of considering $\chi=0$ is that the photons fields are
now totally trapped in their original sheets : photons belonging to
a given sheet are not able to go into the other sheet (as a
noticeable consequence, the structures belonging to a given sheet
are invisible from the perspective of an observer located in the
other sheet). The classical gauge field transformation which reads

\begin{equation}
A_\mu ^{\prime }=A_\mu +\partial _\mu \Lambda  \label{7}
\end{equation}

can be easily extended to fulfil the two-sheeted requirements. A
solution consistent with the above hypothesis $\chi =0$ shows that
the gauge transformation is degenerated and reduced to a single
$e^{iq\Lambda }$ which must be applied to both sheets simultaneously
[30,31]. By setting $\chi =0$ and by considering the same gauge
transformation in the two sheets we can get photons fields
$\mathbf{A}_{\pm }$ which behave independently from each other and
in accordance with observations [30-32]. After introducing the gauge
field into the $Z_2$ Dirac equation and taking the non relativistic
limit (following the standard procedure), a two-sheeted Pauli like
equation can be derived [30-32] (using now natural units)
\begin{equation}
i\hbar \frac \partial {\partial t}\left(
\begin{array}{c}
\left| \psi _{+}\right\rangle \\
\left| \psi _{-}\right\rangle
\end{array}
\right) =\left\{ \mathbf{H}+\mathbf{H}_{cm}(g,\mathbf{A}_{\pm })+\mathbf{H}%
_c(g^2)\right\} \left(
\begin{array}{c}
\left| \psi _{+}\right\rangle \\
\left| \psi _{-}\right\rangle
\end{array}
\right)  \label{8}
\end{equation}
where $\left| \psi _{+}\right\rangle $ and $\left| \psi
_{-}\right\rangle $ correspond to the wave functions in the $(+)$
and $(-)$ sheets respectively. The Hamiltonian $\mathbf{H}$ is a
block-diagonal matrix where each block is simply the classical Pauli
Hamiltonian expressed in both sheets
\begin{equation}
\mathbf{H}_{\pm }=-\frac{\hbar ^2}{2m}\left( \mathbf{\nabla }-i\frac
q\hbar \mathbf{A}_{\pm }\right) ^2+g_s\mu \frac 12\mathbf{\sigma
\cdot B}_{\pm }+V_{\pm }  \label{9}
\end{equation}
and such that $\mathbf{A}_{+}$ and $\mathbf{A}_{-}$ denote the
magnetic vector potentials in the sheet $(+)$ and $(-)$
respectively. The same convention is applied for the magnetic fields
$\mathbf{B}_{\pm }$. $g_s\mu $ is the magnetic moment of the
particle with $g_s$ the gyromagnetic factor and $\mu $ the magneton.
In addition to these ``classical'' terms, the two-sheeted
Hamiltonian comprises also new terms involving the discrete
structure of the bulk. $\mathbf{H}_c$ is a constant geometrical term
proportional to the square of the coupling constant $g^2$ and $\mathbf{H}%
_{cm}$ is another geometrical coupling involving the gauge fields of
the two sheets. This last term is proportional to $g$. Since $g^2$
must be tiny in order for the model to be consistent with the
experimental observations and measurements, the term $\mathbf{H}_c$
can be neglected in a first
approximation [30-32]\footnote{%
The classical finite difference [31] and the non-commutative [30]
approaches of the two-sheeted spacetime predict almost the same
physics. The only difference arises from a difference about the
expression of the $\mathbf{H}_c $ term. In the finite difference
approach an off diagonal part proportional to $g^2 \hbar^2 /m$
appears into $\mathbf{H}_c$. This term was demonstrated to be
responsible of particle oscillations between the two spacetime
sheets, even in the case of a free particle [31]. Nevertheless, as
explained in [31], $\mathbf{H}_c$ can be neglected since it is
proportional to $g^2$, an obvious tiny value in order for the model
to be consistent with
known experimental results. By contrast, in the non-commutative case, the $%
\mathbf{H}_c$ term is free of off diagonal terms, and thus it can be
simply
eliminated through a rescaling of the energy scale.}. As a consequence, $%
\mathbf{H}_{cm}$ imparts most of the new physics of the model. This
contribution takes the form [30-32]
\begin{equation}
\mathbf{H}_{cm}=-ig\gamma g_s\mu \frac 12\left[
\begin{array}{cc}
0 & \mathbf{\sigma \cdot }\left\{ \mathbf{A}_{+}-\mathbf{A}_{-}\right\}  \\
-\mathbf{\sigma \cdot }\left\{ \mathbf{A}_{+}-\mathbf{A}_{-}\right\}
& 0
\end{array}
\right]   \label{10}
\end{equation}
As a consequence, the remaining coupling term $\mathbf{H}_{cm}$
arises through the magnetic vector potentials $\mathbf{A}_{+}$ and
$\mathbf{A}_{-}$
and the magnetic moment $g_s\mu $. In Eq. (10), $\gamma $ is a constant and $%
g_c=\gamma g_s$ stands for the isogyromagnetic factor [31]. From a
theoretical point of view, the neglect of the QED corrections leads to $%
g_s=g_c=2$ for an electron [31]. As a consequence $\gamma $ must be
equal to $1$. Although it was suggested in Ref. [31] that $g_c$
could differ slightly from $g_s$ when taking into account QED, these
corrections will not be considered in the present paper. As a
consequence, in the proton or neutron cases, we consider also
$\gamma =1$ and use the standard experimental values of $g_s$ (i.e.
5.58 and -3.82 respectively) [31]. It is worth being noticed that
$g\left\{ \mathbf{A}_{+}-\mathbf{A}_{-}\right\} $ can be seen as an
extradimensional magnetic field. The similarities between
$\mathbf{H}_{cm}$ and the classical term $g_s\mu (1/2)\mathbf{\sigma
\cdot B}_{\pm }$ become more obvious when considering the presence
of the magnetic moment in Eq. (10).

\section{Dynamics of the particles localization}

The coupling term $\mathbf{H}_{cm}$ arises from the geometrical
proximity between the two sheets. As a consequence, even if these
sheets do not share any 4D link, the particles can possibly move
from one sheet to the other one. Such a delocalization between the
two sheets can be seen as a major drawback of the model but we are
now going to show that environmental interactions constrain the
capability of moving between the sheets. An illustrative example can
be given. Let us consider a fermion moving in a constant magnetic
vector potential $\mathbf{A}_{+}=A\mathbf{e}_z$ created in the first
sheet. Assuming that $\mathbf{H}$ can be expressed in its diagonal
form, the particle dynamics reduces to
\begin{equation}
i\hbar \frac \partial {\partial t}\left| \Psi (t)\right\rangle
=\left\{ \left[
\begin{array}{cc}
E_{+} & 0 \\
0 & E_{-}
\end{array}
\right] \mathbf{+}i\hbar \Omega \left[
\begin{array}{cc}
0 & -\sigma _z \\
\sigma _z & 0
\end{array}
\right] \right\} \left| \Psi (t)\right\rangle   \label{11}
\end{equation}
where $\Omega =gg_s\mu A/(2\hbar )$. Provided that the particle is
originally located in the sheet $(+)$ and unpolarized, it is
straightforward to show that the probability to find the particle in
the $(-)$ sheet is
\begin{equation}
P=\frac 1{1+\zeta ^2}\sin ^2\left( \Omega \sqrt{1+\zeta ^2}t\right)
\label{12}
\end{equation}
with $\zeta =(E_{+}-E_{-})/(2\Omega \hbar ).$ It is worth being
noticed that the difference $\Delta E$ between the eigenvalues
$E_{+}$ and $E_{-}$ of the Hamiltonian $\mathbf{H}$ plays a
fundamental role as it governs the particle ability of reaching the
second sheet (i.e. of moving in the discrete space). For that
reason, it is convenient to recast it in the following form
\begin{equation}
\Delta \mathbf{H}=\Delta V_{grav}+\Delta V_{elec}+\Delta K_A+g_s\mu \frac 12%
\mathbf{\sigma \cdot }\Delta \mathbf{B}  \label{13}
\end{equation}
with
\begin{equation}
K_{A,\pm }=-\frac qm\mathbf{A}_{\pm }\cdot \widehat{\mathbf{P}}+\frac{q^2}{2m%
}\left| \mathbf{A}_{\pm }\right| ^2  \label{14}
\end{equation}
In Eq. (13), the gravitational and electrostatic contributions have
been included in the potential in order to give $\Delta \mathbf{H}$
the most general form. \\In the following, we restrict ourselves to
the case of an irrotational magnetic vector potential such that
$\mathbf{B=0}$. From Eq. (12), it is then clear that if $\zeta =0$,
i.e. $\Delta E=E_{+}-E_{-}=0$, the particle oscillates freely
between the two sheets with a time periodicity $T=2\pi \hbar
/(gg_s\mu A)$. Note that during the oscillations, the particle which
is a $M_4\times Z_2$ entity, behaves sometimes as a four dimensional
$M_4$ one. Indeed, when the maximum of the probability is reached
(i.e. 1), the particle becomes totally localized in a four
dimensional submanifold. From this point of view, it is clear that
the particle is not spontaneaously constrained to stay on a specific
sheet. Such ability to move between the sheets, can be seen as a
discrete counterpart of motions in the bulk sometimes discussed in
brane-world models. \\Although the particles are able to oscillate
between the sheets, four dimensional localization can still occur in
this model. We start by considering $\zeta \neq 0$ i.e. there is now
an effective potential $\Delta \mathbf{H}$ applied onto the
particle. Then, it is trivial to see that the amplitude and the
period of oscillations change. The higher the potential is, the
weaker the oscillation amplitude is. Therefore, in presence of a
strong enough potential, the particle is now confined on its sheet
and cannot reach the other one. Since, the potential is in fact the
sum of all external influences acting on the particle, it becomes
obvious that the confinement arises mainly from environmental
interactions. Such an explanation of the matter confinement in a
multidimensional spacetime is elegant in several aspects: first, it
does not require any ad-hoc bulk field nor any extradimensional
gravitational trapping mechanism [9-23]. Secondly, it is quite
remarkable that the model suggests that the confinement of fermions
may arise from usual 4D interactions (particle scattering, EM
interactions, gravitational fields...). \\Note that the matter
confinement could perhaps be insured even for large value of $g$
provided that the intensity of the environmental interactions are
significant. \\Let us consider more closely the electrostatic and
gravitational contributions in $\Delta E$. With regards to the
electrostatic potentials, the neutrality of matter implies that they
can be neglected at our scale or any larger scale. At smaller scale
however, as inside an atom for instance, the typical potential
energies undergone by an electron are about $-10$ eV, which is a
quite small value. The same cannot be said for the gravitational
potentials. We consider the typical example of a single neutron
interacting with a larger object like the earth, the sun or the
Milky Way core (which is responsible for the sun attraction around
the galactic center). If the acceleration induced by the earth on a
neutron is $a=1$, then it is $a=6\times 10^{-4}$ and $a=2\times
10^{-11}$ for the sun and the Milky Way core respectively. Although
the gravitational forces exerted by a very distant object can be
neglected, the gravitational potentials can become huge as we can
convince oneself easily. For a neutron in the lab reference frame,
the potential energy arising from the earth/particle interaction is
$0.65$ eV (absolute value), this potential reaches $9$ eV when
considering the sun influence and it rises up to $500$ eV for the
Milky Way core. Although the induced acceleration by such a massive
object is negligible in an earth lab, its effect on the potential
energy of particles is considerable. It is of course related to the
fact that the gravitational potential varies like $1/r$ and the
force like $1/r^2$ with $r$ the distance between the particle and
the influencing massive object. In such circumstances, it appears
that the main contribution to the confinement is the gravitational
potential exerted by massive objects or clusters of matter. Until
now we have only discussed the gravitational contribution of the
mass located in our sheet but from Eq. (13), it is clear that the
masses located in the neighboring sheet also constrain the particle
oscillations. Unfortunately, to assess the gravitational influence
of the masses located in the other sheet, we need some kind of
gravitational telescope which does not exist at present time.\\ Let
us now calculate the effective probability for a single particle, to
oscillate between the two sheets. We are going to assume a coupling
constant of the order $g=10^{-1}$ m$^{-1}$ (which is a quite huge
value). Typically,
the magnetic potential can be related to a current intensity $I$ through $%
A=\mu _0I$ ($\mu _0$ is the vacuum permeability). Therefore,
intensity can be considered as a good indicator of the reachable
magnetic potentials. We
thus consider an intensity of $10^9$ A and a global confining potential of $%
500$ eV. For a neutron, the Eq. (12) indicates that the maximum
amplitude of the oscillations of probability is $2.5\cdot 10^{-16}$.
This value is so tiny that it is hard to believe that any particle
oscillations might be observed in the physical world. Hence, in all
practical cases, the confining effect is so strong that the
particles are constrained to move on their sheet only. Such a result
is interesting because it shows that the model does not violate any
current observations. However, it is unsatisfactory because it seems
to prohibit any attempt to confirm the existence of a coupling
between the neighboring sheets as predicted by this model. A closer
investigation of $\mathbf{H}_{cm}$ suggests however a way to enhance
the probability amplitude.

\section{Resonant leap between spacetime sheets}

Obviously, a first way to enhance the probability of oscillations is
to increase the intensity of the vector potential and/or to decrease
the gravitational trapping potential. Unfortunately, since gravity
cannot be shielded and since magnetic potential is limited by the
reachable current intensity, these solutions remain poorly
effective. Previously, the
similarity between $\mathbf{H}_{cm}$ and the term $g_s\mu (1/2)\mathbf{%
\sigma \cdot B}_{\pm }$ was noted. Since this last term is
traditionally involved in magnetic resonance, it is legitimate to
search for a two-sheeted counterpart of this mechanism through which
particle oscillations might occur. Let us again consider the
two-sheeted Pauli equation (Eq. (8)). By analogy with magnetic
resonance, we are now going to consider a particle initially located
in the first sheet and subjected to a rotating curl-free magnetic
vector potential. The irrotational character of the magnetic
potential precludes the existence of a related magnetic field and
thus prevents the existence of a Hamiltonian contribution of the form $g_s\mu (1/2)\mathbf{\sigma \cdot B}$. Let us choose $\mathbf{A}_{+}=\mathbf{%
A}_p=A_p\mathbf{e(}t\mathbf{)}$ and $\mathbf{A}_{-}=\mathbf{0}$ with
the magnetic vector potential rotating in the $Oxy$ plane, i.e.
\begin{equation}
\mathbf{e(}t\mathbf{)=}\left[
\begin{array}{c}
\cos \omega t \\
\sin \omega t \\
0
\end{array}
\right]   \label{15}
\end{equation}
For simplicity reasons, it is assumed that $\mathbf{H}$ can be
expressed in its diagonal form. The eigenstates of the Hamiltonian
$\mathbf{H}$ are defined as
\begin{equation}
\left| \psi _{+1/2}^{+}\right\rangle =\left[
\begin{array}{c}
1 \\
0 \\
0 \\
0
\end{array}
\right] \text{, }\left| \psi _{-1/2}^{+}\right\rangle =\left[
\begin{array}{c}
0 \\
1 \\
0 \\
0
\end{array}
\right]   \label{16}
\end{equation}
for particles located in the first sheet with an energy equals to
$E_{+}$ (assuming both spin states $\pm 1/2$), and
\begin{equation}
\left| \psi _{+1/2}^{-}\right\rangle =\left[
\begin{array}{c}
0 \\
0 \\
1 \\
0
\end{array}
\right] \text{, }\left| \psi _{-1/2}^{-}\right\rangle =\left[
\begin{array}{c}
0 \\
0 \\
0 \\
1
\end{array}
\right]   \label{17}
\end{equation}
for particles located in the second sheet with an energy equals to
$E_{-}$ (assuming both spin states $\pm 1/2$). Taking into account
these assumptions, Eq. (8) becomes
\begin{eqnarray}
i\hbar \frac \partial {\partial t}\left| \Psi (t)\right\rangle &=&
\left\{ \left[
\begin{array}{cc}
E_{+} & 0 \\
0 & E_{-}
\end{array}
\right] +i\Omega _p\hbar \left[
\begin{array}{cc}
0 & -\mathbf{\sigma \cdot e} \\
\mathbf{\sigma \cdot e} & 0
\end{array}
\right] \right\} \left| \Psi (t)\right\rangle   \label{18}
\end{eqnarray}
with $\Omega _p=g\gamma g_s\mu A_p/(2\hbar )=E_p/\hbar $. It is then
possible to develop the general solution as
\begin{eqnarray}
\left| \Psi (t)\right\rangle &=& a(t)\left| \psi
_{+1/2}^{+}\right\rangle e^{-i(1/\hbar )E_{+}t} + b(t)\left| \psi
_{-1/2}^{+}\right\rangle e^{-i(1/\hbar )E_{+}t} \nonumber\\ && {}+
c(t)\left| \psi _{+1/2}^{-}\right\rangle e^{-i(1/\hbar )E_{-}t}
+d(t)\left| \psi _{-1/2}^{-}\right\rangle e^{-i(1/\hbar
)E_{-}t}\label{19}
\end{eqnarray}
This solution can be inserted into Eq. (18) to give the following
system

\begin{eqnarray}
\left\{
\begin{array}{c}
\frac d{dt}a(t)  =-\Omega _pd(t)e^{i(\eta -\omega )t} \\
\frac d{dt}b(t) =-\Omega _pc(t)e^{i(\eta +\omega )t} \\
\frac d{dt}c(t) =\Omega _pb(t)e^{-i(\eta +\omega )t} \\
\frac d{dt}d(t) =\Omega _pa(t)e^{-i(\eta -\omega )t}
\end{array}
\right. \label{20}
\end{eqnarray}

where $\eta =(E_{+}-E_{-})/\hbar $. This system can be trivially
solved by analytical means. Using the initial conditions
$a(t=0)=a_0$, $b(t=0)=b_0$ and $c(t=0)=d(t=0)=0$ with $\left|
a_0\right| ^2+\left| b_0\right| ^2=1$ (i.e. the particle is
initially localized in the $(+)$ sheet), one obtains
\begin{eqnarray}
a(t) &=&a_0 \biggl\{ \cos \left( (1/2)\sqrt{(\eta -\omega )^2+4\Omega _p^2}%
t\right)   \label{21} \\
&& -i\frac{\eta -\omega }{\sqrt{(\eta -\omega )^2+4\Omega _p^2}}\sin
\left( (1/2)\sqrt{(\eta -\omega )^2+4\Omega _p^2}t\right) \biggr\}
e^{i(1/2)(\eta -\omega )t}
 \nonumber \\
b(t) &=&b_0 \biggl\{ \cos \left( (1/2)\sqrt{(\eta +\omega )^2+4\Omega _p^2}%
t\right)   \label{22} \\
&& -i\frac{\eta +\omega }{\sqrt{(\eta +\omega )^2+4\Omega _p^2}}\sin
\left( (1/2)\sqrt{(\eta +\omega )^2+4\Omega _p^2}t\right)
\biggr\}e^{i(1/2)(\eta +\omega )t}
\nonumber \\
c(t) &=&-b_0\frac{2\Omega _p}{\sqrt{(\eta +\omega )^2+4\Omega
_p^2}}\sin \left( (1/2)\sqrt{(\eta +\omega )^2+4\Omega
_p^2}t\right)e^{-i(1/2)(\eta +\omega )t}
\label{23} \\
d(t) &=&-a_0\frac{2\Omega _p}{\sqrt{(\eta -\omega )^2+4\Omega
_p^2}}\sin \left( (1/2)\sqrt{(\eta -\omega )^2+4\Omega
_p^2}t\right)e^{-i(1/2)(\eta -\omega )t} \label{24}
\end{eqnarray}
If the spin was in the $+1/2$ state at $t=0$ ($a_0=1$ and $b_0=0$),
it can then be shown using Eq. (24) that the probability $P$ to find
the particle in the second sheet is given by
\begin{equation}
P=\frac{4\Omega _p^2}{(\eta -\omega )^2+4\Omega _p^2}\sin ^2\left( (1/2)%
\sqrt{(\eta -\omega )^2+4\Omega _p^2}t\right)   \label{25}
\end{equation}
and in addition, the particle is then in the down spin state. By
contrast, if the spin was in the $-1/2$ state at $t=0$ we get for
the probability $P$
\begin{equation}
P=\frac{4\Omega _p^2}{(\eta +\omega )^2+4\Omega _p^2}\sin ^2\left( (1/2)%
\sqrt{(\eta +\omega )^2+4\Omega _p^2}t\right)   \label{26}
\end{equation}
and now the particle is in the spin up state in the second sheet.
Eq. (25) corresponds to a resonant process : as the potential vector
rotates with the angular frequency $\omega =\eta $ the probability
$P$ presents a maximal amplitude equals to one, independently of the
coupling constant and of the magnetic vector potential amplitude. In
addition, the probability oscillates with an angular frequency given
by $\Omega _p$. The resonance width at the half-height is $\Delta
\omega =4\Omega _p$. The weaker the coupling constant $g$ is, the
narrower the resonance is. By contrast, the greater the magnetic
vector potential $A_p$ is, the broader the resonance is. Note that
Eq. (26) is similar to Eq. (25) except that it describes an
anti-resonance associated with a counter-rotating vector potential.
Both Eq. (25) and (26) remind those found in magnetic resonance (MR)
where the spin orientation is influenced by the combination of a
static and an oscillating magnetic field. In the present case, the
matter exchange between the sheets arises from the resonant
interaction between the rotating curl-free magnetic vector potential
and the spin (via the magnetic moment). Note that the diagonal
contribution of the Hamiltonian in Eq. (18) plays a role similar to
the spin/static magnetic field interaction found in MR. But if
$E_{+}-E_{-}$ stands for the difference of magnetic energy between
two spin states in MR, the difference $\Delta E$ in our model arises
from a completely different origin. As a consequence, when the
angular frequency $\omega $ of the magnetic vector potential matches
the confinement potential $\Delta E/\hbar $ (see previous section)
the particle is resonantly transferred from one sheet to the other
one. In practice, any rotating vector potential induces electric
field and possibly magnetic field that can skew dramatically any
experimental study by increasing the contribution of $\Delta E$.
Even if we work with a neutral particle (a neutron for instance),
the use of an irrotational magnetic vector potential is necessary to
avoid any undesirable magnetic field. This requirement may be
difficult to fulfill and an electromagnetic wave is not necessarily
relevant for this purpose. Therefore, obtaining the required
conditions for a resonant transfer remains a major challenge
although it could perhaps be accessible with our current technology.

\section{Conclusion}

In this paper, the quantum behavior of massive particles has been
studied within the framework of the two-sheeted spacetime model
introduced in earlier works. The model predicts that fermions
oscillate between the two spacetime sheets in presence of
irrotational vector potentials. It is shown however that
environmental interactions constrain very tightly these oscillations
and lead to a four-dimensional localization of the fermion species.
Yet, we predict that the environmental confining effect can be
overcome through a resonant mechanism which might be investigated at
a laboratory scale. The study of such a mechanism could be relevant
for demonstrating the existence of another spacetime sheet and
confirms that our spacetime is just a sheet embedded in a more
complex manifold.

\end{document}